\documentclass{acta}
\usepackage{supertabular,lscape,epsfig}
\usepackage{amssymb}
\usepackage{amsmath}
\usepackage{multicol}
\usepackage{graphicx}
\usepackage{longtable}
\usepackage{pdflscape}

\SetPages{0}{0}

\SetVol{58}{2008}

\begin{document}

\begin{Titlepage}
\Title{ Planetary-mass companions to a retired B star BD+37 3172 and a retired F star BD+42 2315.
\footnote{
Based on observations obtained with the Hobby-Eberly Telescope,
which is a joint project of the University of Texas at Austin, the Pennsylvania State University, Stanford University, Ludwig-Maximilians-Universitat M\"unchen, and Georg-August-Universit\"at
G\"ottingen}
\footnote{Based on observations made with the Italian Telescopio Nazionale Galileo (TNG) operated on the island of
 La Palma by the Fundaci\'on Galileo Galilei of the INAF (Istituto Nazionale di Astrofisica) 
 at the Spanish Observatorio del Roque de los Muchachos of the Instituto de Astrof\'isica de Canarias.}}

\Author{Niedzielski$^1$, A., Jaros$^1$, R., Paczuski$^1$, A., Adam\'ow$^2$, M., 
Wolszczan$^{3,4}$, A., Villaver$^{5,6}$, E., Maciejewski$^1$, G., Deka-Szymankiewicz$^1$, B. }
{$^1$Institute of Astronomy, Nicolaus Copernicus University in Toru\'n, ul. Gagarina 11, 87-100 Toru\'n, Poland\\ e-mail:Andrzej.Niedzielski@umk.pl\\
$^2$Center for AstroPhysical Surveys, National Center for Supercomputing Applications, Urbana, IL 61801, USA \\
$^3$Department of Astronomy and Astrophysics, Pennsylvania State University, 525 Davey Laboratory, University Park, PA 16802, USA\\
$^4$Center for Exoplanets and Habitable Worlds, Pennsylvania State University, 525 Davey Laboratory, University Park, PA 16802, USA\\
$^5$Instituto de Astrof\'isica de Canarias, 38205 La Laguna, Tenerife, Spain\\
$^6$Agencia Espacial Espa\~nola (AEE), 41015 Sevilla, Spain
}

\Received{Month Day, Year}
\end{Titlepage}

\Abstract{ 
Radial velocity searches may lead to detection of exoplanets at large orbital separations only if long-enough time-series of data are available. Therefore publication of precise measurements collected in the past is very valuable even if not successfully completed with a definitive detection. 

Here we present 309 precise ($\sigma$RV$\approx$5-7 m s$^{-1}$) multi-epoch radial velocities for 28 stars observed with the Hobby-Eberly Telescope and its High Resolution Spectrograph between 2004 and 2013.

Based on the observations gathered 
we present a low mass companion ($m_{p}\sin{i}$ = 10.6 $M_{J}$ in 1887.76 $\pm$ 0.01 d, 4.65 au orbit with $e$ = 0.59 $\pm$ 0.01) to a K giant BD+37 3172 ($M$=3.75$\pm$ 0.86 M$_{\odot}$), and a planetary mass companion (m$_{p}$sin{\it{i}}=0.55 M$_{J}$  in 123.05$\pm$0.04 day, 0.55 au orbit with e=0.73$\pm$0.03) to a K giant BD+42 2315 ($M$=1.38$\pm$ 0.30 M$_{\odot}$).

We also present  
two preliminary detections of new  spectroscopic binaries: BD+56 2957 (K5) and HD 236555 (G5).
}
{exoplanets; radial velocity technique; stars: intermediate-mass; stars: red giants; stars:binary}


\section{Introduction}

Over 30 years after the discovery of the first planets orbiting other stars (Wolszczan \& Frail 1992; Mayor \& Queloz 1995), it is clear that exoplanets are 
common. With over 5600 exoplanets known today around low-mass (0.08-2.2 M$_{\odot}$) stars, we have reasons to assume that planets accompany most if not all stars in the Galaxy (e.g., Batalha et al. 2013, 
Fressin et al. 2013, Poleski et al. 2021).

Most of our knowledge about exoplanets results from massive radial velocity (RV) or transit searches focused on solar-mass stars (Winn, 2018).
Recently, an abundant population of planets orbiting low-mass stars, M dwarfs, was discovered and intensively studied (Dressing \& Charbonneau 2015,   
Gaidos 2016). 

At the same time, the population of exoplanets orbiting the intermediate-mass (2.2-8 M$_{\odot}$) stars remains somewhat elusive. 
Of $\approx$5600 exoplanet hosts with known masses in an exoplanet.eu, only 43 are intermediate-mass stars, and only 10 with masses above 3 M$_{\odot}$.

The most efficient way to detect exoplanets around such stars is to apply the RV method to evolved intermediate-mass stars on the giant branch after they cooled down and spun down past the MS evolution and, in consequence, developed rich spectra suitable for the Doppler technique. Such studies (see  Ottoni et al. 2022 for a complete list of active projects), however, result typically in the detection of planets around solar-mass stars again, and statistical considerations show the occurrence rate to quickly decrease beyond $\approx$2M$_{\odot}$  (Reffert et al. 2015; Wolthoff et al. 2022). 
Similar findings are delivered by  Kepler and TESS missions that also detected planets around subgiants and red giant branch stars (e.g. Lillo-Box et al. 2014; Grunblatt et al. 2022).

These results find support in theoretical studies (Kennedy \& Kenyon 2008) as intense stellar irradiation overcomes accretion hindering the formation of planetary cores. On the other hand, intermediate-mass stars' protoplanetary disks are expected to be massive, which suggests that the planet population around these stars might be rich. These planets are thought to be formed beyond the snow-line, and they subsequently migrate inwards in the disc (e.g. Mordasini et al. 2009). In the case of intermediate-mass stars theory predicts (Kennedy \& Kenyon 2008) that the gradual evolution of the protoplanetary disc causes migration to slow down. As a consequence, planets orbiting intermediate-mass stars may be absent in close-in orbits (observable with the RV technique) but abundant at larger orbital separations (observable with the direct imaging technique). 

The Atacama Large Millimeter/submillimeter Array (ALMA) observations of protoplanetary disks around young intermediate-mass Herbig stars (Stapper et al. 2022) indeed show that these disks are on average larger and have higher dust masses  compared to the less-massive pre-main sequence T Tauri stars'  disks.

This line of reasoning is supported by the first outcome of the BEAST survey (Janson et al. 2021a, b; Viswanath et al. 2023)
in which the direct imaging technique is applied to B-type stars (M$\ge$ 2.5 M$_{\odot}$). It shows that sub-stellar companions can form around this kind of star.
The intermediate-mass star exoplanet population apparently exists but has not been studied well so far. 

One of the obstacles is obviously the expected long orbital period which, 
together with a low value of the planet to stellar radius ratio, makes spotting of such planets through transits unlikely. 
Detection of exoplanets around the intermediate-mass stars with the RV technique requires very long observing runs (see for instance Delgado Mena et al. 2023). It is therefore desirable to make archive precise RV measurements of such objects available to quicken detection in this yet not well-known parameter space.  

Here we present precise, multi-epoch RV measurements obtained with the HET/HRS for a sample of stars 
that contains, among the others,  8 intermediate-mass stars   
(TYC 0405 00684 1,
HD 187552,
TYC 2818 00602 1, 
TYC 2818 00990 1,
BD+37 3172,
HD 16992, 
HIP 46275, 
HD 236555
).
In the case of  BD+37 3172 we present evidence of a low-mass companion.

\section{The sample, observations and data reduction}

The Pennsylvania-Toru\'n Planety Search (PTPS) was initiated in 2004 with a sample of about 600 giants from the Red Giant Clump (Niedzielski$\&$ Wolszczan 2008a) observed with the Hobby-Eberly Telescope (HET, Ramsey et al. 1998). 
The sample was swiftly extended with another $\approx$400 evolved stars (Niedzielski $\&$ Wolszczan 2008b, c), and the total sample of $\approx$1000 GK-giants brighter than 11 mag, occupying the
area in the HR-diagram, which is approximately defined by the MS, the instability strip,
and the coronal dividing line, was monitored for RV variations with HET. In a series of papers, (Zieli\'nski et al. 2012, Adam\'ow et al. 2014, Adamczyk et al. 2016, Niedzielski et al. 2016, Deka-Szymankiewicz et al. 2018) detailed spectroscopic analysis of the observed stars was presented, resulting in atmospheric parameter ($T_{\mathrm{eff}}$, logg, [Fe/H], v$_{rot}$ sin{\it{i}}, basic abundances including Li) determinations as well as integral parameter (M/M$_{\odot}$, R/R$_{\odot}$, logL/L$_{\odot}$, age) estimates.
In the turn of spectroscopic analysis, the complete sample of 885 PTPS stars was defined (Deka-Szymankiewicz et al. 2018, hereafter BDS), composed of 132 dwarfs, 238 subgiants, and 515 giants.  Numerous targets initially observed with HET were excluded for various reasons: lack of adequate or high enough quality data at the time of sample definition, detected variations, and inconsistencies in derived parameters (see Niedzielski et al. 2016 and BDS for details).
Here, we present precise radial velocities for the complete sample of 28 stars initially observed within the PTPS, with multiple high-quality epochs of RV measurements available but not included in the final sample.

Observations presented here were made with
HET 
equipped with the High Resolution Spectrograph (HRS, Tull 1998) in the queue scheduled mode (Shetrone et al.
2007). The instrumental configuration and observing procedure were identical to those described
by Cochran et al. (2004). The spectrograph, fed with
the 2 arcsec fiber, was used in the R = 60,000 resolution
mode with a gas cell (I2) inserted into the optical path.

The HET/HRS is a general-purpose spectrograph,  neither temperature nor pressure-controlled. Therefore the calibration
of the RV measurements with this instrument is best accomplished
with the I2 gas cell technique (Marcy \& Butler
1992; Butler et al. 1996). Our application of this technique
to HET/HRS data is described in detail in Nowak (2012) and
Nowak et al. (2013). With the typical RV precision levels of a few m s$^{-1}$ 
we use the Stumpff (1980) algorithm to refer the measured RVs to
the Solar System barycenter.

The HET/HRS spectral line bisector (BIS) 
measurements were obtained from the spectra used for the
I2 gas-cell technique (Marcy \& Butler 1992; Butler et al. 1996).
The combined stellar and iodine spectra were first cleaned of
the I2 lines by dividing them by the corresponding iodine spectra
imprinted in the flat-field spectra and then cross-correlated
with a binary K2 star mask. A detailed description of this procedure
is described in Nowak et al. (2013). 

The BIS measures the asymmetry
of a spectral line, which can arise for multiple  reasons like the blending of
lines or a stellar surface feature  (see Gray 2005 for a discussion of BIS properties).
The BIS has proven to be a powerful tool to detect starspots
and background binaries (Queloz et al. 2001; Santos et al. 2002)
that can mimic a planet signal in the RV data  
(see Santerne et al. 2015
and G\"unther et al. 2018 for an extensive discussion). Unfortunately, for the
slow-rotating giant stars like many of our targets, the BIS is not a very sensitive
activity indicator (Santos et al. 2003, 2014).

For BD+37 3172 we also obtained 8  epochs of RV  data with the 3.58 meter Telescopio Nazionale Galileo (TNG)
and its High Accuracy Radial velocity Planet Searcher in the North hemisphere
(Harps-N, Cosentino et al. 2012).
The RVs were calculated by cross-correlating the stellar spectra with the digital mask for a K2-type star. The TNG/Harps-N spectra were processed with the standard user's pipeline, Data Reduction Software (DRS; Pepe et al. 2002; Lovis \& Pepe 2007).

The complete list of targets and their basic atmospheric parameters are presented in Table 1. This table also contains information about 
the time span of observations ($\Delta$T),
the number of available epochs of observations (N$_{obs}$),  
observed amplitude in radial velocities ($\Delta$RV), 
the mean RV uncertainty ($\sigma_{RV}$),
the values of the correlation coefficient between the radial velocities and spectral line bisector ($r$), and the corresponding $p$-value ($p$).

\section{Results}

In Table 2 we present 309 epochs of radial velocities collected for the 28 targets presented in Table 1. In Table 1 the stellar effective temperature $T_{eff}$, logarithm of the stellar surface gravitational acceleration log$g$ [cgs] and the stellar metallicity [Fe/H] were taken from Zieli\'nski $et al.$ (2012) and Adamczyk $et al.$ (2016). 

\begin{landscape}
\MakeTable{llllrrrrrrrrr}{12.5cm}{A summary of basic data on target stars}
{\hline
Ident & Ident & Teff& logg  & [Fe/H] &  $\Delta$T    & N$_{obs}$   & $\Delta$RV & $\sigma_{RV}$  &  r  & p \\
TYC   & other & [K] & [cgs] &        & [day]         &  & [m s$^{-1}$] & [m s$^{-1}$]   &      & \\
\hline
1 & 2 & 3 & 4 & 5 & 6 & 7 & 8 & 9 & 10 & 11 \\
\hline
0405 00684 1 &  & 4750 $\pm$ 250 & 1.50 $\pm$ 0.50 & -0.11 $\pm$ 0.14 & 2542 & 5 & 115 & 5.3 & 0.14 & 0.83 \\
0938 01112 1 & HIP 77226 & 5190 $\pm$ 20 & 4.28 $\pm$ 0.06 & -0.82 $\pm$ 0.02 & 1805 & 4 & 46 & 8.6 & 0.68 & 0.32 \\
1062 00017 1 & HD 187552 & 5000 $\pm$ 250 & 2.00 $\pm$ 0.50 & -0.07 $\pm$ 0.13 & 1110 & 4 & 39 & 4.4 & 0.48 & 0.52 \\
1430 00351 1 & HIP 54459 & 4860 $\pm$ 35 & 4.12 $\pm$ 0.14 & -0.64 $\pm$ 0.03 & 2264 & 28 & 80 & 7.8 & -0.04 & 0.83 \\
1496 00374 1 & HIP 78047 & 4500 $\pm$ 250 & 2.50 $\pm$ 0.50 & 0.00 $\pm$ 0.23 & 651 & 5 & 74 & 5.9 & -0.62 & 0.26 \\
1814 01132 1 & HIP 18995 & 5566 $\pm$ 25 & 2.28 $\pm$ 0.05 & -1.16 $\pm$ 0.03 & 2092 & 6 & 170 & 15.0 & -0.15 & 0.78 \\
2809 00167 1 & HIP 3669 & 4935 $\pm$ 25 & 4.29 $\pm$ 0.09 & -0.43 $\pm$ 0.02 & 2600 & 10 & 861 & 9.0 & -0.73 & 0.02 \\
2818 00602 1 &  & 4750 $\pm$ 250 & 1.50 $\pm$ 0.50 & -0.45 $\pm$ 0.17 & 1828 & 6 & 140 & 7.9 & -0.32 & 0.54 \\
2818 00990 1 &  & 5250 $\pm$ 250 & 1.50 $\pm$ 0.50 & -0.07 $\pm$ 0.12 & 128 & 3 & 21 & 7.5 & -0.28 & 0.82 \\
3020 01288 1 & BD+42 2315 & 4500 $\pm$ 250 & 2.00 $\pm$ 0.50 & -0.17 $\pm$ 0.14 & 3331 & 43 & 97 & 6.5 & 0.20 & 0.21 \\
3105 01103 1 & BD+37 3172 & 4500 $\pm$ 250 & 1.50 $\pm$ 0.50 & -0.03 $\pm$ 0.16 & 3096 & 26 & 179 & 5.5 & 0.47 & 0.02 \\
3226 00868 1 & HIP 112047 & 4250 $\pm$ 250 & 3.00 $\pm$ 0.50 & -0.51 $\pm$ 0.73 & 2626 & 15 & 21413 & 8.7 & 0.39 & 0.15 \\
3304 00479 1 & HD 16992 & 4750 $\pm$ 250 & 1.50 $\pm$ 0.50 & -0.18 $\pm$ 0.12 & 2072 & 8 & 45 & 5.3 & -0.61 & 0.11 \\
3431 00086 1 &  & 4750 $\pm$ 250 & 1.50 $\pm$ 0.50 & -0.27 $\pm$ 0.26 & 1110 & 3 & 14 & 7.8 & 0.99 & 0.07 \\
3431 01221 1 & HIP 46275 & 5021 $\pm$ 20 & 4.15 $\pm$ 0.06 & -0.13 $\pm$ 0.03 & 1109 & 3 & 25 & 6.6 & -0.96 & 0.18 \\
3435 00030 1 & BD+46 1608 & 5168 $\pm$ 28 & 3.70 $\pm$ 0.09 & -0.72 $\pm$ 0.03 & 1030 & 2 & 8554 & 15.7 & 1.00 & 1.00 \\
3463 01145 1 & HIP 66262 & 4691 $\pm$ 20 & 4.08 $\pm$ 0.07 & -0.39 $\pm$ 0.03 & 3022 & 4 & 6712 & 7.2 & -0.66 & 0.34 \\
3486 00642 1 & HIP 76929 & 5216 $\pm$ 25 & 4.38 $\pm$ 0.09 & -0.42 $\pm$ 0.02 & 62 & 2 & 22 & 7.4 & 1.00 & 1.00 \\
3498 00634 1 & HIP 79941 & 4731 $\pm$ 5 & 3.92 $\pm$ 0.02 & -0.39 $\pm$ 0.02 & 2748 & 19 & 26 & 6.0 & 0.20 & 0.40 \\
3663 00838 1 & HD 236555& 5000 $\pm$ 250 & 2.00 $\pm$ 0.50 & -0.06 $\pm$ 0.30 & 2552 & 17 & 503 & 11.1 & -0.21 & 0.42 \\
3819 01043 1 & HD 237903 & 4398 $\pm$ 92 & 3.45 $\pm$ 0.37 & -0.70 $\pm$ 0.09 & 277 & 5 & 12 & 4.6 & -0.51 & 0.38 \\
3826 00664 1 & HIP 52668 & 5329 $\pm$ 20 & 4.27 $\pm$ 0.06 & -0.14 $\pm$ 0.03 & 3016 & 14 & 3929 & 12.2 & 0.11 & 0.70 \\
3947 02317 1 & HIP 98123 & 4851 $\pm$ 12 & 3.74 $\pm$ 0.05 & -0.79 $\pm$ 0.04 & 1648 & 2 & 315 & 6.2 & -1.00 & 1.00 \\
4006 00890 1 & BD+56 2957 & 4000 $\pm$ 250 & 2.00 $\pm$ 0.50 & 0.21 $\pm$ 0.21 & 2962 & 9 & 10330 & 7.2 & 0.43 & 0.25 \\
4099 01786 1 & HIP 29548 & 4752 $\pm$ 112 & 4.05 $\pm$ 0.50 & -0.31 $\pm$ 0.09 & 3032 & 60 & 65 & 6.4 & 0.06 & 0.63 \\
4820 03585 1 & HIP 35173 & 4842 $\pm$ 8 & 4.30 $\pm$ 0.03 & 0.13 $\pm$ 0.01 & 30 & 2 & 14 & 4.1 & 1.00 & 1.00 \\
4835 00774 1 & HD 61606B & 4796 $\pm$ 10 & 3.76 $\pm$ 0.04 & -0.62 $\pm$ 0.03 & 649 & 4 & 22 & 6.4 & 0.33 & 0.67 \\
\hline
\vspace{-55.91882pt}
}

\end{landscape}
In the last two columns, Pearsons correlation coefficient between RV and BIS (r) and the corresponding p value (statistical significance) are presented. 

For every epoch, the table contains target identification (TYC), the epoch of observations (MJD=JD-2400000.5), as well as the value of radial velocity, its uncertainty, spectral line bisector, and its uncertainty, all values in m s$^{-1}$.

As the targets presented here were not selected for the final PTPS sample, their observations ceased, and in many cases, a limited number of epochs are available, which allows for preliminary, putative detections of companions only. However, the high precision of the gathered radial velocity measurements makes them suitable for other, more in-depth studies. Below, we present the most interesting cases that certainly deserve more attention.

For seven stars 
(BD+63 639,
G 202-38,
BD+42 2315,
BD+43 4270,
BD+16 2216,
HD 236555,  
BD+37 3172)
we collected 15 epochs of radial velocities or more, enough for a preliminary search for periodic signals (Figure 1). We identified such signals only in the case of BD+37 3172  and BD+42 2315. 

%

%

For these two stars, we 
completed Keplerian analysis using  a
hybrid approach (e.g., Go\'zdziewski et al. 2003; Go\'zdziewski and Migaszewski, 2006;
Go\'zdziewski et al. 2007), that combines the PIKAIA-based, global genetic algorithm (GA; Charbonneau 1995) with a faster and more precise local method. The
best-fit Keplerian orbit was delivered by RVLIN (Wright and Howard 2009) within a narrow parameter range found by the GA 
semi-global search. The uncertainties were
estimated with the bootstrap method described by Marcy et al. (2005). Niedzielski
et al. (2015a) provide a more detailed description of the Keplerian analysis.

The red giants are known to present various types of variability (see, for instance, Niedzielski et al. (2021) for an extensive discussion); therefore, for the most promising detections presented here, we  also checked for alternatives to Keplerian scenarios.

\subsection{BD+37 3172  an intermediate-mass star with a planetary-mass companion}

BD+37 3172 is an evolved thick galactic disk giant with T$_{eff}$= 4500$\pm$250 K; [Fe/H]=-0.03$\pm$0.16; logg=1.5$\pm$0.5 (Adamczyk et al. 2015), and v{\it{sini}}= 1.8$\pm$0.7 km s$^{-1}$ (Adam\'ow et al. 2014). According to  Adamczyk et al. (2015) it's mass, radius and luminosity are: 
M/M$_{\odot}$=3.75$\pm$ 0.86, 
R/R$_{\odot}$=50.24$\pm$28.97,  
log(L/L$_{\odot}$)=2.84$\pm$0.28. 
This star was not included in the PTPS complete sample defined in BDS because its atmospheric parameters in Zieli\'nski et al. (2012) were unavailable and were obtained with a slightly different method by Adamczyk et al. (2015).

We collected for this star 26 epochs of HET/HRS data over 3096 days ($\approx$8.5 yr) that show an amplitude of 179 m s$^{-1}$ and average uncertainty of 5.5m s$^{-1}$. We also gathered additional 8 epochs of TNG/Harps-N data for the star over a period of 1056 days, that show an amplitude of 169 m s$^{-1}$ and mean uncertainty of 2.2 m s$^{-1}$ (Table 3).


\begin{table}
\TabCap{10cm}{Multiepoch radial velocities and spectral line bisectors for 28 target stars}
\centering
\TableFont
\begin{tabular}{lccccc}

\hline
Ident& MJD & RV         & $\sigma {RV}$ & BIS  & $\sigma {BIS}$\\
TYC  &     & [ms$^{-1}$] &  [ms$^{-1}$]     & [ms$^{-1}$] &  [ms$^{-1}$]\\

\hline
0405 00684 1 &  53928.27027 &     -78.55 &       4.27 &       9.14 &      13.30 \\
0405 00684 1 &  55031.13255 &      21.43 &       6.15 &       7.08 &       9.58 \\
0405 00684 1 &  55751.14778 &      35.06 &       5.00 &     -10.19 &      13.05 \\
0405 00684 1 &  56439.28711 &      36.86 &       5.80 &      34.52 &      17.99 \\
0405 00684 1 &  56470.18951 &      36.86 &       5.42 &      19.87 &      17.49 \\
0938 01112 1 &  53545.27028 &      -2.62 &       6.80 &    -166.75 &      21.87 \\
0938 01112 1 &  53546.28136 &     -25.08 &       9.33 &    -127.45 &      20.98 \\
0938 01112 1 &  53941.19185 &      20.84 &       9.54 &     -52.87 &      20.71 \\
0938 01112 1 &  55350.34111 &       8.98 &       8.80 &     -48.31 &      19.45 \\
1062 00017 1 &  53921.25595 &       2.09 &       3.52 &      73.48 &       8.84 \\
1062 00017 1 &  53926.22716 &     -11.29 &       3.72 &      57.83 &       8.99 \\
1062 00017 1 &  54929.48505 &      27.56 &       5.31 &      65.45 &      12.62 \\
1062 00017 1 &  55031.20449 &      -7.76 &       4.85 &      34.01 &       7.57 \\
1430 00351 1 &  53423.41626 &       9.73 &       7.89 &     -50.47 &      39.73 \\
1430 00351 1 &  53687.50211 &       5.80 &       5.81 &    -105.04 &      24.93 \\
1430 00351 1 &  53694.48512 &      -1.89 &       6.10 &    -130.06 &      25.38 \\
1430 00351 1 &  53695.48066 &      -2.64 &       6.16 &    -104.36 &      24.14 \\
1430 00351 1 &  54110.36079 &     -24.59 &       7.95 &    -121.37 &      36.70 \\
1430 00351 1 &  54130.48811 &       0.57 &       6.73 &     -37.27 &      23.67 \\
1430 00351 1 &  54167.19856 &       4.97 &       6.02 &    -145.02 &      26.19 \\
1430 00351 1 &  54452.42096 &     -18.20 &       6.88 &     -87.38 &      37.55 \\
1430 00351 1 &  54507.28002 &      20.67 &       7.67 &    -105.18 &      36.72 \\
1430 00351 1 &  54552.14292 &      23.00 &       7.46 &     -79.68 &      28.56 \\
1430 00351 1 &  54848.53344 &       3.21 &       8.19 &     -82.53 &      15.90 \\
1430 00351 1 &  54874.25712 &      31.87 &       7.80 &    -177.43 &      36.58 \\
1430 00351 1 &  54876.25383 &      29.60 &       6.77 &     -96.72 &      29.72 \\
1430 00351 1 &  55171.45629 &      -0.65 &       5.85 &    -109.14 &      26.24 \\
1430 00351 1 &  55218.31080 &     -18.10 &       7.74 &     -70.86 &      32.55 \\
1430 00351 1 &  55240.24013 &       1.15 &       8.57 &    -189.08 &      44.28 \\
1430 00351 1 &  55252.21537 &       2.23 &       8.05 &    -119.49 &      31.48 \\
1430 00351 1 &  55326.21340 &     -21.16 &      11.71 &     -86.72 &      56.93 \\
1430 00351 1 &  55510.51721 &     -22.92 &       7.54 &     -66.47 &      39.64 \\
1430 00351 1 &  55557.38351 &      11.41 &       9.86 &    -135.28 &      43.35 \\
1430 00351 1 &  55558.37876 &     -20.93 &       7.54 &    -145.42 &      35.69 \\
1430 00351 1 &  55561.39824 &       1.08 &      13.03 &    -261.52 &      65.51 \\
1430 00351 1 &  55566.37344 &     -48.25 &       7.88 &    -138.80 &      42.83 \\
1430 00351 1 &  55606.25241 &      -0.08 &       6.94 &    -163.41 &      28.45 \\
1430 00351 1 &  55646.33133 &      11.91 &       8.69 &     -96.10 &      34.02 \\
1430 00351 1 &  55653.11779 &       6.32 &       6.82 &     -89.79 &      20.92 \\
1430 00351 1 &  55674.26649 &      -1.08 &       6.98 &    -119.36 &      26.65 \\
1430 00351 1 &  55687.24198 &       1.32 &       9.85 &    -118.04 &      41.40 \\
1496 00374 1 &  54614.35631 &      13.10 &       4.96 &      -6.91 &      13.10 \\
1496 00374 1 &  54614.37161 &      14.43 &       5.14 &      -8.61 &      12.48 \\
1496 00374 1 &  54640.29525 &      12.29 &       5.86 &       8.36 &      15.29 \\
\hline
\end{tabular}
\end{table}

\setcounter{table}{1}
\begin{table}

\TabCap{10cm}{Continued}
\centering
\TableFont
\begin{tabular}{lccccc}

\hline
Ident& MJD & RV         & $\sigma {RV}$ & BIS  & $\sigma {BIS}$\\
TYC  &     & [ms$^{-1}$] &  [ms$^{-1}$]     & [ms$^{-1}$] &  [ms$^{-1}$]\\

\hline
1496 00374 1 &  54929.31307 &       2.16 &       7.20 &     -39.30 &      21.33 \\
1496 00374 1 &  55265.39277 &     -59.36 &       6.34 &      28.91 &       9.01 \\
1814 01132 1 &  53762.20157 &       8.57 &      18.22 &     143.50 &      72.41 \\
1814 01132 1 &  53800.09874 &      49.14 &      13.94 &      50.19 &     158.36 \\
1814 01132 1 &  54107.27928 &      -4.97 &      17.39 &     323.73 &     143.20 \\
1814 01132 1 &  54173.09576 &     -11.85 &      10.90 &     250.80 &     119.05 \\
1814 01132 1 &  55635.10365 &    -109.48 &      15.32 &     118.96 &     154.64 \\
1814 01132 1 &  55854.25970 &      60.61 &      14.12 &     132.49 &     148.89 \\
2809 00167 1 &  53632.42830 &     485.40 &      12.41 &     -56.38 &      40.22 \\
2809 00167 1 &  54365.44580 &     273.49 &       7.00 &      45.85 &      18.49 \\
2809 00167 1 &  54470.13494 &     252.25 &       8.86 &     -27.82 &      19.17 \\
2809 00167 1 &  54640.44030 &     234.12 &       9.04 &       1.91 &      24.50 \\
2809 00167 1 &  55488.34000 &     -87.86 &       6.85 &     -15.21 &      20.42 \\
2809 00167 1 &  55518.28263 &     -95.50 &       8.17 &      27.78 &      23.55 \\
2809 00167 1 &  55818.20352 &    -129.63 &      11.47 &     110.48 &      40.87 \\
2809 00167 1 &  55840.39009 &    -220.60 &       9.31 &      38.86 &      22.00 \\
2809 00167 1 &  55875.30923 &    -222.61 &       8.22 &      42.23 &      20.32 \\
2809 00167 1 &  56232.32854 &    -375.78 &       8.96 &     112.29 &      18.88 \\
2818 00602 1 &  54403.35392 &     -28.16 &       7.41 &     -28.03 &      13.38 \\
2818 00602 1 &  55474.41605 &       9.83 &       8.17 &      20.49 &      35.68 \\
2818 00602 1 &  55519.31065 &     -90.47 &       9.26 &     -21.96 &      25.04 \\
2818 00602 1 &  55854.38453 &      49.95 &       7.47 &     -47.41 &      19.22 \\
2818 00602 1 &  55884.10154 &      37.71 &       7.57 &     -61.56 &      22.88 \\
2818 00602 1 &  56231.35869 &      -7.66 &       7.61 &      -3.76 &      18.78 \\
2818 00990 1 &  54658.41741 &      14.41 &       9.78 &      14.31 &      25.45 \\
2818 00990 1 &  54785.33491 &      -6.73 &       7.02 &      15.30 &      19.35 \\
2818 00990 1 &  54786.31472 &      -0.47 &       5.76 &      31.12 &      13.41 \\
3020 01288 1 &  53024.36249 &      41.91 &      10.36 &      34.64 &      20.43 \\
3020 01288 1 &  53094.15196 &     -15.93 &       7.61 &      24.93 &      19.48 \\
3020 01288 1 &  53156.23666 &     -26.98 &       5.72 &       7.09 &       9.33 \\
3020 01288 1 &  53430.46427 &      28.02 &       7.97 &      37.23 &      19.92 \\
3020 01288 1 &  53804.22622 &     -15.88 &       6.07 &      27.97 &      17.56 \\
3020 01288 1 &  54201.13211 &      -5.33 &       6.43 &      29.27 &      20.86 \\
3020 01288 1 &  54253.24202 &      16.55 &       6.56 &      25.23 &      15.72 \\
3020 01288 1 &  54843.39459 &      -1.18 &       7.69 &      64.54 &       9.10 \\
3020 01288 1 &  54845.37399 &      35.86 &       7.72 &      58.47 &      20.76 \\
3020 01288 1 &  54861.34574 &       3.03 &       7.62 &      26.14 &      20.85 \\
3020 01288 1 &  54876.29499 &      10.93 &       6.97 &       4.51 &      16.01 \\
3020 01288 1 &  54881.51192 &     -35.20 &       7.63 &     -19.98 &      16.62 \\
3020 01288 1 &  55022.14431 &      -3.57 &       6.00 &      17.16 &       9.39 \\
3020 01288 1 &  55252.26755 &     -22.25 &       6.52 &      49.00 &      16.94 \\
3020 01288 1 &  55264.46341 &     -24.18 &       5.60 &      47.11 &      12.03 \\
3020 01288 1 &  55308.33270 &      10.88 &       6.29 &     -12.62 &      14.56 \\
3020 01288 1 &  55371.16652 &     -35.20 &       6.01 &     -28.27 &      13.81 \\
\hline
\end{tabular}
\end{table}

\newpage
\setcounter{table}{1}
\MakeTable{lccccc}{10cm}{Continued}
{\hline
Ident& MJD & RV         & $\sigma {RV}$ & BIS  & $\sigma {BIS}$\\
TYC  &     & [ms$^{-1}$] &  [ms$^{-1}$]     & [ms$^{-1}$] &  [ms$^{-1}$]\\

\hline

3020 01288 1 &  55582.35736 &      32.05 &       6.65 &     -12.25 &      18.00 \\
3020 01288 1 &  55591.34786 &      17.68 &       7.53 &      29.22 &      19.34 \\
3020 01288 1 &  55605.29381 &      38.77 &       6.64 &      28.43 &      18.86 \\
3020 01288 1 &  55610.50248 &      40.37 &       5.12 &      -2.54 &      10.65 \\
3020 01288 1 &  55615.49743 &      -7.27 &       6.40 &     -31.06 &      12.27 \\
3020 01288 1 &  55620.48426 &     -15.45 &       5.88 &      18.81 &      20.26 \\
3020 01288 1 &  55626.48182 &     -20.07 &       7.26 &      16.57 &      20.57 \\
3020 01288 1 &  55632.22940 &     -19.08 &       6.01 &      27.95 &      15.23 \\
3020 01288 1 &  55637.20120 &     -38.11 &       5.93 &      18.36 &      16.68 \\
3020 01288 1 &  55643.20107 &      -7.02 &       7.56 &      -0.06 &      15.71 \\
3020 01288 1 &  55685.30913 &      52.07 &       6.37 &      21.90 &      15.22 \\
3020 01288 1 &  55702.26279 &      26.99 &       5.88 &     -18.53 &      11.33 \\
3020 01288 1 &  55709.24257 &      12.54 &       5.70 &      20.21 &      12.94 \\
3020 01288 1 &  55723.20799 &      -5.51 &       5.72 &      39.20 &      12.94 \\
3020 01288 1 &  55729.18140 &      -1.34 &       5.78 &      64.71 &      15.35 \\
3020 01288 1 &  55743.14125 &     -26.09 &       6.31 &       2.37 &      12.42 \\
3020 01288 1 &  55749.13964 &     -44.50 &       5.90 &      -1.64 &      12.33 \\
3020 01288 1 &  55964.31641 &      13.46 &       6.14 &      35.31 &      12.50 \\
3020 01288 1 &  56093.18956 &      31.02 &       6.86 &       5.59 &      18.43 \\
3020 01288 1 &  56102.16615 &       3.62 &       5.63 &      -9.66 &      13.35 \\
3020 01288 1 &  56122.12501 &      -5.04 &       7.08 &       0.03 &      17.99 \\
3020 01288 1 &  56256.50153 &     -20.18 &       6.86 &     -18.36 &      16.89 \\
3020 01288 1 &  56259.48550 &      11.72 &       5.67 &       5.79 &      14.89 \\
3020 01288 1 &  56308.37338 &      36.24 &       6.16 &      66.85 &      15.18 \\
3020 01288 1 &  56323.32534 &       3.53 &       6.16 &      28.97 &      18.66 \\
3020 01288 1 &  56355.23566 &     -21.67 &       5.37 &      36.67 &      14.59 \\
3105 01103 1 &  53307.06938 &     -23.19 &       6.21 &      10.35 &      21.65 \\
3105 01103 1 &  53307.07540 &     -22.69 &       8.57 &       1.33 &      31.09 \\
3105 01103 1 &  53546.41351 &     -19.21 &       8.17 &      20.28 &      17.63 \\
3105 01103 1 &  53895.22719 &      40.12 &       4.41 &      44.38 &      10.03 \\
3105 01103 1 &  53932.37278 &      56.73 &       4.95 &      -6.75 &       9.05 \\
3105 01103 1 &  54047.05088 &      15.60 &       4.14 &      24.70 &      11.00 \\
3105 01103 1 &  54191.41026 &      45.62 &       4.94 &      17.66 &      11.11 \\
3105 01103 1 &  54546.43437 &     -82.70 &       6.21 &     -41.96 &      20.09 \\
3105 01103 1 &  55312.34795 &     -53.83 &       4.12 &      37.57 &      13.60 \\
3105 01103 1 &  55362.20809 &     -28.49 &       5.15 &      29.72 &      12.53 \\
3105 01103 1 &  55390.12818 &     -28.65 &       7.28 &      46.40 &      18.01 \\
3105 01103 1 &  55444.22551 &      -9.47 &       5.01 &      -4.02 &      12.54 \\
3105 01103 1 &  55483.09541 &      -4.40 &       5.29 &       5.78 &      12.99 \\
3105 01103 1 &  55510.05049 &     -18.00 &       5.06 &     -12.73 &      13.49 \\
3105 01103 1 &  55616.52450 &      20.77 &       5.28 &      45.44 &      16.20 \\
3105 01103 1 &  55639.45354 &      23.49 &       5.79 &      12.98 &      20.91 \\
3105 01103 1 &  55640.46198 &      25.06 &       4.65 &      22.08 &      15.48 \\
3105 01103 1 &  55651.39990 &      14.61 &       6.08 &     -19.24 &      19.26 \\
\hline
}

\newpage
\setcounter{table}{1}
\MakeTable{lccccc}{10cm}{Continued}
{\hline
Ident& MJD & RV         & $\sigma {RV}$ & BIS  & $\sigma {BIS}$\\
TYC  &     & [ms$^{-1}$] &  [ms$^{-1}$]     & [ms$^{-1}$] &  [ms$^{-1}$]\\

\hline

3105 01103 1 &  55704.26920 &       8.49 &       5.23 &      60.08 &      13.23 \\
3105 01103 1 &  55738.18564 &      35.31 &       5.28 &      -3.12 &      13.66 \\
3105 01103 1 &  55775.30091 &      50.20 &       5.28 &      61.19 &      11.58 \\
3105 01103 1 &  55807.22893 &      59.08 &       5.43 &      15.50 &      11.00 \\
3105 01103 1 &  55849.11323 &      60.36 &       4.73 &      21.69 &       9.64 \\
3105 01103 1 &  56203.15654 &     -64.04 &       4.96 &      13.63 &      10.88 \\
3105 01103 1 &  56371.44341 &    -119.07 &       6.14 &     -26.67 &      21.28 \\
3105 01103 1 &  56403.35450 &     -87.47 &       5.30 &     -37.88 &      13.90 \\
3226 00868 1 &  53664.26165 &   -8971.04 &       8.56 &     242.59 &      28.79 \\
3226 00868 1 &  53681.19666 &   -4747.28 &       7.94 &     237.83 &      32.90 \\
3226 00868 1 &  53682.20907 &   -4629.65 &       7.73 &     233.13 &      31.32 \\
3226 00868 1 &  55776.23605 &    4753.11 &       8.10 &     275.73 &      31.56 \\
3226 00868 1 &  55779.22376 &    5181.83 &       9.12 &     377.61 &      56.83 \\
3226 00868 1 &  55795.20130 &    6584.57 &       9.76 &     392.82 &      48.65 \\
3226 00868 1 &  55815.14604 &    7524.96 &      10.81 &     724.70 &      64.45 \\
3226 00868 1 &  55830.34335 &    7191.58 &       7.03 &     340.76 &      51.73 \\
3226 00868 1 &  55842.08031 &    5733.70 &       8.52 &     261.94 &      52.10 \\
3226 00868 1 &  55893.15700 &   -2257.82 &       8.18 &     395.82 &      26.44 \\
3226 00868 1 &  55926.09000 &   -9279.13 &       8.77 &     236.00 &      34.17 \\
3226 00868 1 &  56097.35684 &  -13888.86 &       9.78 &     143.58 &      40.50 \\
3226 00868 1 &  56149.46950 &   -1306.04 &       8.78 &     147.40 &      22.71 \\
3226 00868 1 &  56204.09196 &    6975.65 &       8.85 &      24.21 &      46.22 \\
3226 00868 1 &  56290.08819 &     691.55 &       8.21 &     309.06 &      28.45 \\
3304 00479 1 &  53372.21050 &       0.92 &       6.15 &      46.72 &      17.56 \\
3304 00479 1 &  53372.21736 &      -5.67 &       5.44 &      37.49 &      12.95 \\
3304 00479 1 &  53680.15328 &       4.24 &       4.06 &      31.01 &      12.46 \\
3304 00479 1 &  53683.37188 &      -2.94 &       4.10 &      35.85 &      11.58 \\
3304 00479 1 &  53686.14132 &       5.54 &       4.98 &      35.60 &      15.33 \\
3304 00479 1 &  54007.49121 &     -15.06 &       4.73 &      18.86 &       8.05 \\
3304 00479 1 &  54446.09476 &      29.87 &       6.61 &      -3.24 &      18.53 \\
3304 00479 1 &  55444.36104 &      -5.30 &       5.97 &      53.01 &      15.17 \\
3431 00086 1 &  54785.42377 &      -8.79 &       7.58 &      24.63 &      22.23 \\
3431 00086 1 &  54856.42721 &       2.80 &       9.06 &      67.83 &      19.73 \\
3431 00086 1 &  55895.36130 &       5.36 &       6.72 &      71.10 &      17.10 \\
3431 01221 1 &  53451.28258 &       3.62 &       6.42 &     -44.37 &       8.45 \\
3431 01221 1 &  54463.51009 &     -12.81 &       6.31 &     -40.14 &      15.77 \\
3431 01221 1 &  54560.23032 &      12.01 &       7.17 &     -50.25 &      25.94 \\
3435 00030 1 &  53455.32092 &    3675.95 &      14.64 &     615.55 &      34.43 \\
3435 00030 1 &  54485.49412 &   -4878.60 &      16.86 &     -53.19 &      47.19 \\
3463 01145 1 &  53451.24459 &   -4195.11 &       8.02 &     -43.44 &      11.88 \\
3463 01145 1 &  54218.14938 &    -753.69 &       8.04 &    -111.24 &      24.02 \\
3463 01145 1 &  56432.30486 &    2516.63 &       6.20 &    -104.15 &      23.85 \\
3463 01145 1 &  56473.18646 &     503.30 &       6.67 &    -168.67 &      27.76 \\
3486 00642 1 &  54225.23209 &     -11.14 &       7.45 &    -121.44 &      29.87 \\

\hline
}

\newpage
\setcounter{table}{1}
\MakeTable{lccccc}{10cm}{Continued}
{\hline
Ident& MJD & RV         & $\sigma {RV}$ & BIS  & $\sigma {BIS}$\\
TYC  &     & [ms$^{-1}$] &  [ms$^{-1}$]     & [ms$^{-1}$] &  [ms$^{-1}$]\\

\hline

3486 00642 1 &  54287.26244 &      11.14 &       7.45 &    -112.12 &      12.94 \\
3498 00634 1 &  53424.43216 &      -6.81 &       7.17 &     -90.52 &      29.84 \\
3498 00634 1 &  53601.15182 &      -8.73 &       5.42 &     -96.18 &      21.54 \\
3498 00634 1 &  53927.25341 &       2.33 &       5.71 &    -138.91 &      20.26 \\
3498 00634 1 &  54253.37624 &      -0.10 &       5.98 &     -92.58 &      19.56 \\
3498 00634 1 &  54287.28429 &       1.10 &       5.62 &    -111.89 &      13.32 \\
3498 00634 1 &  54339.13218 &     -10.01 &       5.59 &    -112.13 &      13.53 \\
3498 00634 1 &  54564.31453 &      10.15 &       7.17 &     -83.02 &      26.35 \\
3498 00634 1 &  54604.42425 &       9.41 &       6.05 &    -108.63 &      20.18 \\
3498 00634 1 &  54726.10256 &      -2.89 &       6.65 &    -119.38 &      26.21 \\
3498 00634 1 &  55252.43031 &       5.55 &       6.63 &     -71.68 &      26.60 \\
3498 00634 1 &  55259.40327 &      15.14 &       5.87 &     -98.51 &      22.62 \\
3498 00634 1 &  55264.38189 &       0.77 &       5.93 &    -145.86 &      24.09 \\
3498 00634 1 &  55265.37929 &       9.03 &       5.33 &     -86.36 &      20.33 \\
3498 00634 1 &  55276.36057 &       3.85 &       6.70 &    -118.10 &      21.79 \\
3498 00634 1 &  55347.18437 &     -11.21 &       5.61 &     -91.71 &      18.03 \\
3498 00634 1 &  55392.27143 &      -6.46 &       5.93 &    -140.17 &      22.07 \\
3498 00634 1 &  55582.51070 &      -0.52 &       6.44 &    -101.01 &      26.63 \\
3498 00634 1 &  55621.42175 &       2.20 &       5.15 &    -107.38 &      22.12 \\
3498 00634 1 &  56172.12329 &      -8.24 &       5.58 &    -113.14 &      18.54 \\
3663 00838 1 &  53688.25389 &     390.23 &      10.90 &    -112.73 &      42.57 \\
3663 00838 1 &  54081.22212 &     213.53 &       8.31 &    -113.24 &      27.91 \\
3663 00838 1 &  54786.27366 &      30.39 &       9.08 &     -94.39 &      31.52 \\
3663 00838 1 &  54792.26815 &      42.28 &      10.66 &    -182.37 &      34.70 \\
3663 00838 1 &  55179.18969 &     -57.00 &      19.72 &     -15.87 &      67.26 \\
3663 00838 1 &  55227.06788 &     -86.01 &      13.66 &     -55.92 &      42.93 \\
3663 00838 1 &  55227.07623 &     -90.73 &      11.42 &     -35.04 &      30.69 \\
3663 00838 1 &  55481.36734 &     -87.93 &      14.64 &     -53.38 &      27.50 \\
3663 00838 1 &  55511.26815 &     -75.10 &      10.94 &    -184.73 &      30.98 \\
3663 00838 1 &  55514.27494 &     -54.59 &      11.14 &    -132.53 &      31.07 \\
3663 00838 1 &  55755.39811 &     -86.31 &      11.30 &    -100.81 &      28.07 \\
3663 00838 1 &  55801.46957 &     -95.28 &       8.99 &    -102.72 &      25.00 \\
3663 00838 1 &  55888.03458 &     -38.65 &       8.51 &    -168.26 &      24.94 \\
3663 00838 1 &  55931.14929 &    -113.07 &      10.64 &     -86.23 &      35.53 \\
3663 00838 1 &  56205.37218 &     -57.75 &      10.29 &     -93.28 &      29.42 \\
3663 00838 1 &  56233.11193 &     -23.42 &       9.70 &     -28.71 &      35.91 \\
3663 00838 1 &  56240.28655 &     -10.38 &       8.04 &     -94.77 &      26.76 \\
3819 01043 1 &  54872.43486 &      -1.49 &       6.25 &    -128.07 &      11.95 \\
3819 01043 1 &  55138.49966 &       4.45 &       3.40 &    -116.85 &      17.38 \\
3819 01043 1 &  55145.48797 &       5.12 &       4.20 &     -95.10 &      26.54 \\
3819 01043 1 &  55146.49245 &      -6.79 &       5.35 &     -48.23 &      32.68 \\
3819 01043 1 &  55149.48978 &      -5.31 &       3.64 &    -102.08 &      17.38 \\
3826 00664 1 &  53395.28903 &    2695.04 &      12.91 &      16.04 &      12.91 \\
3826 00664 1 &  54130.28649 &    1674.76 &      11.06 &      57.33 &      22.77 \\

\hline
}
\newpage
\setcounter{table}{1}
\MakeTable{lccccc}{10cm}{Continued}
{\hline
Ident& MJD & RV         & $\sigma {RV}$ & BIS  & $\sigma {BIS}$\\
TYC  &     & [ms$^{-1}$] &  [ms$^{-1}$]     & [ms$^{-1}$] &  [ms$^{-1}$]\\

\hline

3826 00664 1 &  54166.18804 &    1582.25 &      11.15 &      82.15 &      23.36 \\
3826 00664 1 &  54908.34127 &     546.43 &      11.73 &     231.90 &      29.12 \\
3826 00664 1 &  54922.32579 &     508.93 &      13.34 &     245.16 &      16.19 \\
3826 00664 1 &  55510.49112 &    -382.35 &      12.00 &      85.57 &      43.25 \\
3826 00664 1 &  55627.18631 &    -525.94 &      12.62 &      98.50 &      38.01 \\
3826 00664 1 &  55647.32557 &    -545.30 &      11.09 &      71.67 &      25.29 \\
3826 00664 1 &  55875.50155 &    -834.36 &      12.14 &       4.45 &      24.80 \\
3826 00664 1 &  55917.39707 &    -812.05 &      11.79 &      86.50 &      28.63 \\
3826 00664 1 &  55956.44667 &    -879.29 &      12.37 &     111.79 &      35.17 \\
3826 00664 1 &  56308.34951 &   -1159.36 &      13.27 &      28.22 &      34.77 \\
3826 00664 1 &  56353.37115 &   -1200.99 &      12.71 &      83.67 &      31.84 \\
3826 00664 1 &  56411.22969 &   -1233.83 &      12.61 &       5.28 &      34.19 \\
3947 02317 1 &  54599.36686 &     153.25 &       6.09 &    -127.02 &      47.62 \\
3947 02317 1 &  56247.04810 &    -161.80 &       6.25 &     -55.91 &      35.26 \\
4006 00890 1 &  53543.42763 &    3088.50 &       7.08 &     -28.53 &      21.61 \\
4006 00890 1 &  53694.17958 &    2589.72 &       5.92 &     -16.95 &      20.88 \\
4006 00890 1 &  53905.44299 &    1784.31 &       5.73 &     -38.41 &      15.89 \\
4006 00890 1 &  54049.18698 &    1044.25 &       5.54 &      21.42 &      17.98 \\
4006 00890 1 &  54069.14733 &     982.99 &       5.65 &      -7.00 &      17.24 \\
4006 00890 1 &  54399.07367 &    -966.01 &       6.32 &      14.68 &      17.11 \\
4006 00890 1 &  54453.12657 &   -1379.12 &      12.80 &    -133.19 &      28.60 \\
4006 00890 1 &  56453.45691 &   -7241.97 &       9.10 &    -107.57 &      53.08 \\
4006 00890 1 &  56505.30178 &   -6079.62 &       6.72 &     -19.86 &      26.51 \\
4099 01786 1 &  53340.42804 &      14.76 &       7.93 &    -108.63 &      29.51 \\
4099 01786 1 &  53422.20469 &       7.11 &       5.66 &    -101.54 &      25.17 \\
4099 01786 1 &  53686.46057 &       8.75 &       5.37 &    -116.00 &      17.84 \\
4099 01786 1 &  53687.44815 &       1.14 &       5.82 &     -89.69 &      17.29 \\
4099 01786 1 &  53689.44540 &      -1.07 &       5.69 &    -104.35 &      17.52 \\
4099 01786 1 &  53694.45065 &       9.82 &       5.23 &     -95.14 &      17.39 \\
4099 01786 1 &  53696.31799 &      14.50 &       5.16 &     -89.84 &      14.34 \\
4099 01786 1 &  53697.45828 &      12.25 &       5.28 &    -109.62 &      16.35 \\
4099 01786 1 &  54397.39713 &      -5.24 &       6.69 &    -106.75 &      11.29 \\
4099 01786 1 &  54421.43643 &       2.65 &       5.57 &     -77.62 &      20.44 \\
4099 01786 1 &  54437.41295 &      13.18 &       5.95 &    -124.44 &      16.02 \\
4099 01786 1 &  54483.27307 &      22.10 &       5.98 &     -87.64 &      21.45 \\
4099 01786 1 &  54525.19836 &      11.58 &       6.47 &    -116.00 &      19.90 \\
4099 01786 1 &  54545.12630 &      11.35 &       6.46 &     -93.19 &      17.52 \\
4099 01786 1 &  54843.32358 &      10.94 &       6.82 &     -87.73 &      12.31 \\
4099 01786 1 &  54850.18184 &      13.50 &       5.41 &    -109.66 &      21.91 \\
4099 01786 1 &  54856.26739 &       4.04 &       6.73 &    -144.88 &      24.21 \\
4099 01786 1 &  54861.12533 &      14.35 &       7.92 &    -165.64 &      28.80 \\
4099 01786 1 &  54868.11100 &      14.17 &       7.94 &    -107.19 &      29.74 \\
4099 01786 1 &  54874.11080 &      11.29 &       7.60 &    -117.86 &      24.84 \\
4099 01786 1 &  55106.46692 &      16.07 &       6.18 &     -88.09 &      11.60 \\
4099 01786 1 &  55110.46675 &      -1.40 &       7.01 &    -132.91 &      11.52 \\

\hline
\vspace{-21.57419pt}
}

\newpage
\setcounter{table}{1}
\MakeTable{lccccc}{10cm}{Concluded}
{\hline
Ident& MJD & RV         & $\sigma {RV}$ & BIS  & $\sigma {BIS}$\\
TYC  &     & [ms$^{-1}$] &  [ms$^{-1}$]     & [ms$^{-1}$] &  [ms$^{-1}$]\\

\hline

4099 01786 1 &  55173.28117 &     -12.51 &       6.66 &     -95.81 &      21.82 \\
4099 01786 1 &  55198.20743 &      -4.60 &       5.94 &    -144.70 &      19.63 \\
4099 01786 1 &  55223.13152 &     -12.73 &       7.06 &    -139.07 &      24.18 \\
4099 01786 1 &  55247.19174 &      -1.78 &       7.10 &    -135.88 &      30.57 \\
4099 01786 1 &  55476.47038 &      -6.29 &       5.21 &     -86.74 &      16.50 \\
4099 01786 1 &  55498.40237 &     -43.15 &       7.60 &     -48.60 &      24.52 \\
4099 01786 1 &  55513.37142 &      -2.90 &       6.53 &     -88.26 &      17.95 \\
4099 01786 1 &  55523.45826 &     -25.70 &       5.70 &     -73.23 &      23.61 \\
4099 01786 1 &  55527.42855 &     -34.69 &       5.87 &    -160.43 &      18.21 \\
4099 01786 1 &  55527.45045 &     -34.29 &       5.84 &    -118.04 &      19.46 \\
4099 01786 1 &  55542.35937 &       8.46 &       7.70 &     -78.50 &      32.35 \\
4099 01786 1 &  55563.20064 &      -6.70 &       5.32 &    -136.34 &      19.12 \\
4099 01786 1 &  55567.28601 &       9.92 &       7.93 &     -88.63 &      33.27 \\
4099 01786 1 &  55570.33582 &     -13.27 &       8.07 &    -160.80 &      33.47 \\
4099 01786 1 &  55576.17481 &      -3.40 &       6.06 &    -157.55 &      24.08 \\
4099 01786 1 &  55577.16719 &       2.96 &       6.04 &    -111.79 &      18.52 \\
4099 01786 1 &  55578.18710 &      -1.19 &       5.29 &     -97.65 &      17.31 \\
4099 01786 1 &  55579.17949 &       0.96 &       9.74 &     -76.74 &      26.15 \\
4099 01786 1 &  55580.14398 &      -2.40 &       6.24 &    -106.84 &      26.61 \\
4099 01786 1 &  55581.15593 &      -0.19 &       5.86 &    -107.63 &      23.50 \\
4099 01786 1 &  55581.25477 &       0.34 &       5.51 &     -99.01 &      19.68 \\
4099 01786 1 &  55582.24349 &     -10.18 &       5.82 &    -101.85 &      22.47 \\
4099 01786 1 &  55584.27028 &     -31.60 &       6.80 &    -128.80 &      19.38 \\
4099 01786 1 &  55828.46981 &       3.21 &       5.64 &     -87.12 &      14.84 \\
4099 01786 1 &  55846.43538 &      -6.42 &       7.14 &    -104.64 &      27.57 \\
4099 01786 1 &  55860.49113 &       3.98 &       5.65 &    -118.00 &      14.74 \\
4099 01786 1 &  55888.29446 &      -7.16 &       5.23 &    -108.55 &      15.50 \\
4099 01786 1 &  55916.24735 &       1.45 &       5.13 &    -128.78 &      17.33 \\
4099 01786 1 &  55944.13464 &       4.73 &       6.81 &     -29.31 &      33.12 \\
4099 01786 1 &  55973.09131 &      -2.69 &       5.17 &     -85.79 &      15.05 \\
4099 01786 1 &  56206.40962 &     -26.94 &       7.97 &     -90.37 &      36.62 \\
4099 01786 1 &  56285.33848 &      -1.26 &       5.11 &    -104.94 &      18.61 \\
4099 01786 1 &  56306.20113 &       2.83 &       6.59 &    -130.00 &      26.51 \\
4099 01786 1 &  56318.27395 &      10.96 &       6.62 &    -116.66 &      27.62 \\
4099 01786 1 &  56320.23763 &       6.88 &       6.97 &     -89.41 &      26.61 \\
4099 01786 1 &  56345.21149 &       5.31 &       7.76 &    -108.45 &      33.09 \\
4099 01786 1 &  56359.12867 &       9.31 &       6.00 &     -96.45 &      19.95 \\
4099 01786 1 &  56372.11681 &       3.93 &       5.97 &     -81.40 &      20.16 \\
4820 03585 1 &  54889.17983 &       7.35 &       4.11 &     -25.47 &      11.59 \\
4820 03585 1 &  54919.10951 &      -7.13 &       4.05 &     -76.48 &       7.98 \\
4835 00774 1 &  54853.29954 &      13.79 &       6.71 &     -35.03 &      12.37 \\
4835 00774 1 &  54869.24937 &       1.61 &       5.75 &     -22.98 &      12.48 \\
4835 00774 1 &  54881.17498 &      -8.40 &       6.62 &     -28.96 &      21.46 \\
4835 00774 1 &  55502.46059 &      -7.04 &       6.55 &     -70.88 &      22.07 \\

\hline
}

\clearpage
\newpage 
\begin{figure}[htb!]
\includegraphics[width=1.0\textwidth]{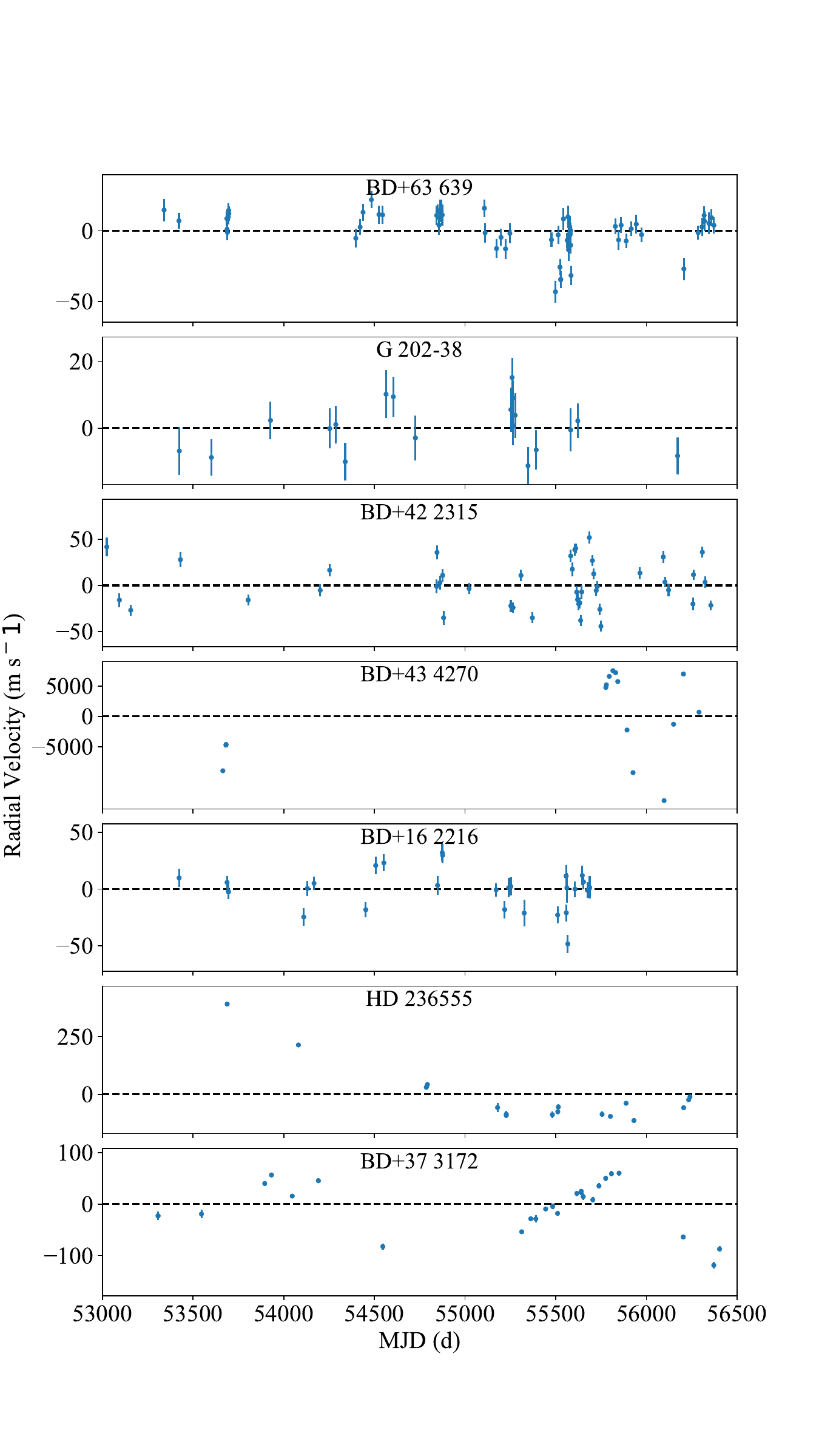}
\FigCap{Radial velocities for the seven stars with 15 or more epochs of observations.}
\vspace{-171.2843pt}
\end{figure}

\clearpage
\newpage
The observed RV amplitude is $\approx$4 times greater than the estimated RV amplitude of p-mode oscillations (k$_{osc}$=43 m s$^{-1}$) (Kjeldsen \& Bedding 1995). P-mode oscillations, with typical periods (in red giants)  of hours to days are not resolved in our low cadence observations and contribute to radial velocity uncertainties only.

\MakeTable{llrr}{12.5cm}{TNG/ HARPS-N data for BD+37 3172}
{\hline
MJD&   RV         & $\sigma_{RV}$ & BVS  \\
  &  [m s$^{-1}$] &  [m s$^{-1}$]     & [m s$^{-1}$] \\
\hline
57135.18548   &   -52562.41  &  2.62  &   105.93       \\   
57168.09973   &   -52558.17  &  1.27  &   101.45        \\  
57196.13539   &   -52552.22  &  1.78  &    89.94       \\   
57237.98979   &   -52517.69  &  1.94  &    97.66       \\    
57238.04161   &   -52530.24  &  1.84  &    100.45      \\    
58044.89203   &   -52547.82  &  4.34  &     87.85      \\    
58072.82486   &   -52577.41  &  0.89  &     99.61      \\   
58191.22804   &   -52687.45  &  2.77  &    100.37     \\

\hline

}

The Pearson's correlation coefficient between RV and BIS (HET) is r=0.47, and the value of the confidence level  p=0.02 makes is statistically insignificant. Also, the Harps-N data show no statistically significant correlation of line bisector with RV  with r=-0.21 (HET/HRS and Harps-N spectral line bisectors are defined in a different way and cannot be directly compared). 
The assumption that the observed RV shifts are due to the Doppler effect is therefore justified. 
The available RV data suggest a long period of $\approx$1800 days.

Our Keplerian analysis of the detected signal reveals a low-mass, m$_{2}$ sin{\it i}=10.6 M$_{J}$ companion on an eccentric (e=0.59$\pm$0.01) orbit with a semi-major axis of 4.65 au (1887.76$\pm$0.01 day). The semi amplitude K=89.94$\pm$0.01 is $\approx$ 16 larger than HET/HRS data uncertainly. The orbital period found is $\approx$2.5 time shorter than the total time span of our data (4884 days). 

The results of our Keplerian analysis are presented in Table 4 and in Figure 2.

%
%

\MakeTable{lll}{12.5cm}{Results of Keplerian analysis for BD+37 3172 and BD+42 2315.}
{\hline
Parameter & TYC 3105 01103 1 & TYC 3020 01288 1 \\
 & BD+37 3172 & BD+42 2315 \\
\hline
P(days) & 1887.76 $\pm$ 0.01            & 123.05 $\pm$ 0.04\\
T$_{0}$ (MJD) & 56191.034 $\pm$ 0.01    & 53155.16 $\pm$ 18.94\\
K (m s$^{-1}$) & 89.94 $\pm$ 0.01        & 26.7 $\pm$ 1.6\\
e & 0.59 $\pm$ 0.01                     & 0.73 $\pm$ 0.03\\
$\omega$ (deg) & 100.07 $\pm$ 0.01      & 109.2 $\pm$ 5.9 \\
m$_{2}$ sin{\it{i}} (m$_{J}$)& 10.6       & 0.55 \\
a (au) & 4.65                           & 0.54 \\
V$_{0}$ (m s$^{-1}$) & -52562.29 $\pm$ 0.01 & \\
offset (m s$^{-1}$)  &52535.1108         & \\
$\sqrt{\chi^{2}}$ & 1.58                & 7.04\\
rms$_{HET}$ (m s$^{-1}$) & 15.27         & 16.35\\
rms$_{TNG}$ (m s$^{-1}$) & 17.83         & \\
N$_{obs}$ &   34                           & 43 \\
\hline

}

\begin{figure}[htb]
\includegraphics[width=1.0\textwidth]{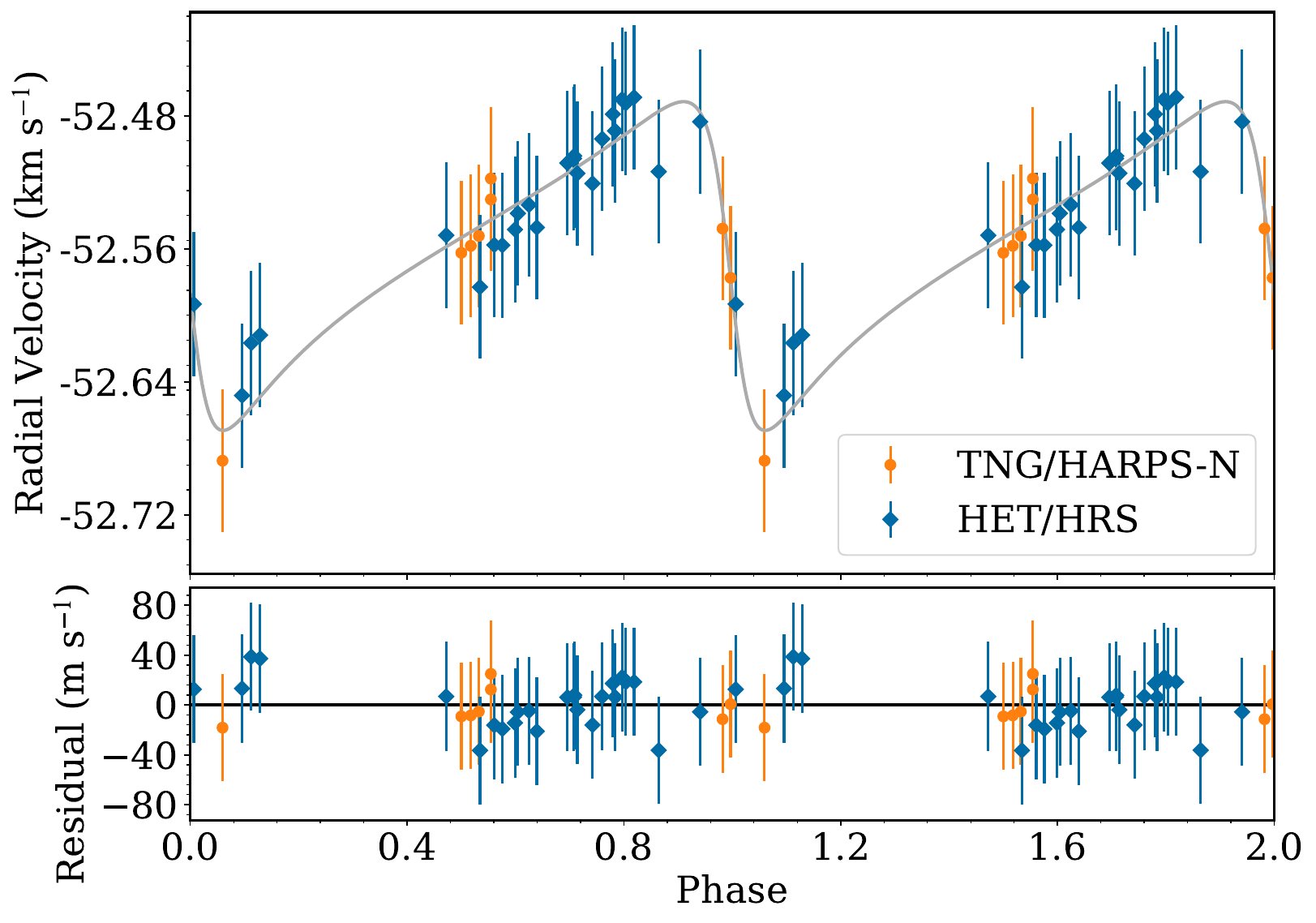}
\FigCap{Best orbital fit for BD+37 3172 b.  The estimated radial velocity scatter due to p-mode oscillations was added to uncertainties. The orbital phase coverage is doubled for better visualization.}
\end{figure}

Given the rotational period estimate v{\it{sini}}= 1.8$\pm$0.7 km s$^{-1}$ and the estimated radius of R/R$_{\odot}$=50.24$\pm$28.97 the face value of the maximum rotational period of BD+37 3172 is $\approx$1406 days. This estimate is rather uncertain given large relative uncertainties in radius and rotational velocity. We can state that within the uncertainties we cannot exclude that the  observed radial velocity period, as shorter than maximum estimated rotational period,  may be induced by a surface feature rotating with the star. We can also conclude that BD+37 3172 is not a fast rotator.

\noindent For BD+37 3172 nearly 9.000 epochs of photometric observations were gathered within SWASP (Pollacco et al. 2006) over $\approx$1547 days between HJD=2453141.625 and HJD=2454688.565, distributed in three observing seasons. These data show an average apparent brightness of 9.53 mag with a mean uncertainty of 0.02 mag and a standard deviation of 0.01 mag. No statistically significant periodicity is present in these data. 

The object was also observed within ASAS (Pojma\'nski 1997) where 273 epochs of observations were gathered over $\sim$1679 days between JD=2456735.065 and JD=2458413.698, and distributed roughly uniformly over the observational time-span. These data show an average apparent magnitude of 9.19 mag with an average uncertainty of 0.02 mag and a standard deviation of 0.06. No periodicity is present in the data, but a long-term trend, longer than the time-span of the observations, is apparent.

Existing photometric observations do not point to any periodic variability of the star similar to the detected RV signal.

Assuming on the other hand, that a hypothetical spot on the surface of BD+37 3172  covers a fraction of the stellar disc similar to the mean uncertainty of existing photometric data (f=0.02) we can estimate after Hatzes (2002) the expected amplitude of observer RV variations as $\approx$ 14 m s$^{-1}$, which is much less than the observed RV semi-amplitude K=89.94$\pm$0.01. This estimate, together with the lack of statistically significant correlation between RV and BIS as well as non-zero eccentricity, makes the spot hypothesis unlikely.

The only interpretation of the observed RV variations in BD+37 3172 supported by existing data is, therefore, its periodic movement around the barycenter due to a low-mass companion.


\subsection{BD+42 2315 - a red giant planetary mass companion host candidate}

BD+42 2315 is a thin galactic disk evolved giant with T$_{eff}$= 4500$\pm$250 K; [Fe/H]=-0.2$\pm$0.11; logg=2$\pm$0.5 (Adamczyk et al. 2015)
and 
v{\it{sini}}= 2.2$\pm$0.5 km s$^{-1}$ (Adam\'ow et al. 2014).
The integrated parameters of this star were estimated in 
Adamczyk et al. (2015): 
M/M$_{\odot}$=1.38$\pm$ 0.30, 
R/R$_{\odot}$=17.85$\pm$10.01, 
log(L/L$_{\odot}$)=1.99$\pm$0.26.
BD+42 2315 was not included in the PTPS complete sample defined in BDS because its atmospheric parameters in Zieli\'nski et al. (2012) were lacking and were obtained later on with a slightly different method by Adamczyk et al. (2015).
For this star we obtained 43 epochs of HET/HRS radial velocities over a period of 3331 days with an average uncertainty of 6.5 m s$^{-1}$. These data show an RV amplitude of 97 m s$^{-1}$ and no correlation with spectral line bisector (r=0.20, p=0.21). The available RV shows a periodic signal of $\approx$120 days. The  RV amplitude is $\approx$6 times greater than the estimated RV amplitude of p-mode oscillations (k$_{osc}$=17 m s$^{-1}$).

Keplerian analysis shows that  
BD+42 2315 hosts an m$_{2}$ sin{\it i}=0.55 m$_{J}$ companion on a very eccentric orbit (e=0.73$\pm$0.03) 
with a semi-major axis of
0.54 au.
The orbital period is much shorter than the time-span of our observations.
The semi amplitude of K=26.7$\pm$1.6 is only 4 times larger than HET/HRS uncertainty. 
The results of our Keplerian analysis are presented in Table 4 and in Figure 3.

\begin{figure}[htb]
\includegraphics[width=1.0\textwidth]{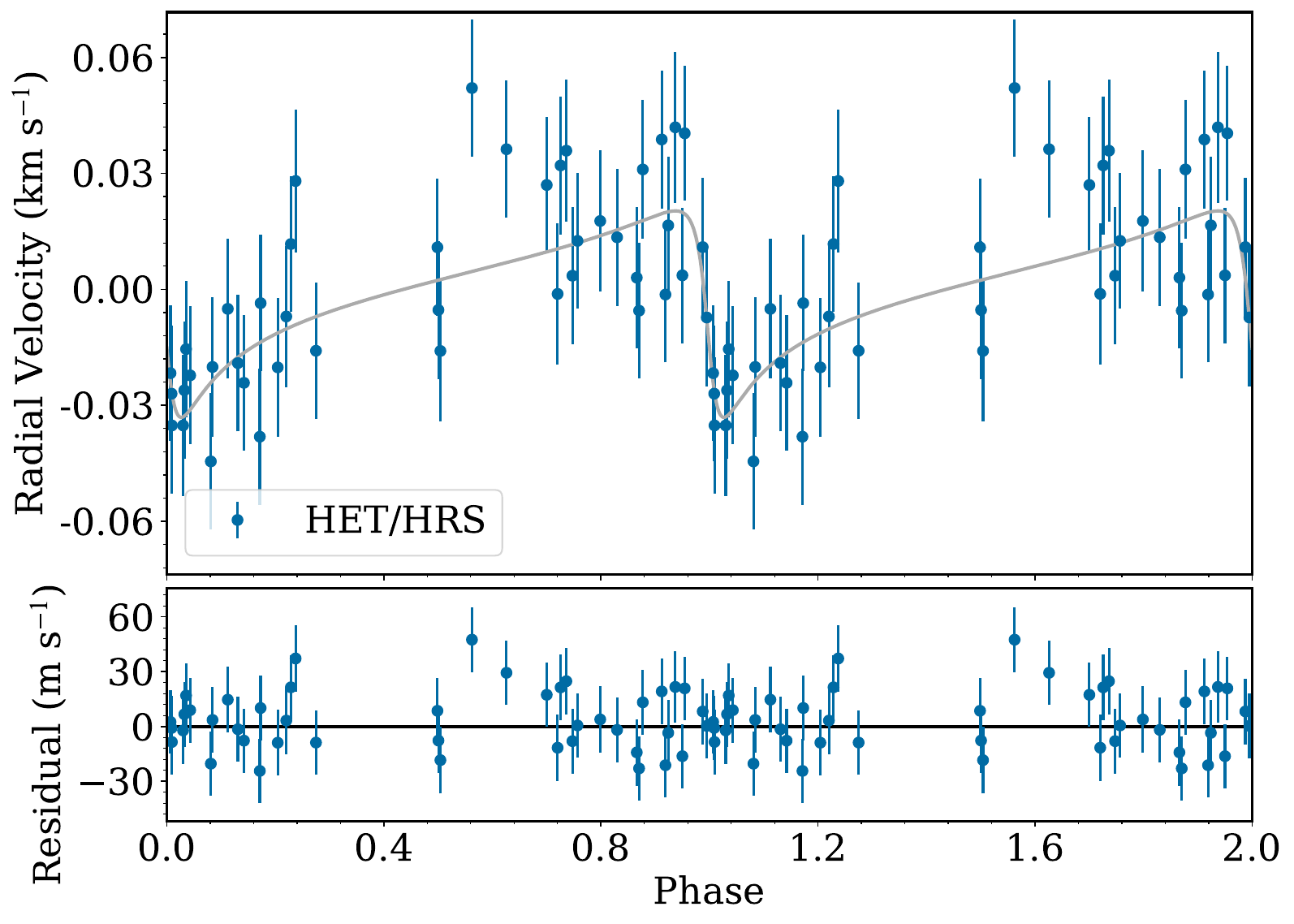}
\FigCap{Best orbital fit for BD+42 2315 b.  The estimated radial velocity scatter due to p-mode oscillations was added to uncertainties. The orbital phase coverage is doubled for better visualization.}
\end{figure}


The estimated rotational velocity of v{\it{sini}}= 2.2$\pm$0.5 km s$^{-1}$ and radius R/R$_{\odot}$ = 17.85$\pm$10.01 point to a maximum rotational velocity of (face value) $\approx$ 414 d. This value is much greater than the observed RV variations period, so the observed RV variations may very well be induced by a spot rotating with the star.

For BD+42 2315 nearly 7000 epochs of photometric observations gathered over 1476 days between HJD=2453128.383 and HJD=2454604.422 are available from SWASP. These data, gathered in three observational seasons show an average apparent brightness of 9.92 mag, an average uncertainty of 0.02 mag and a standard deviation of 0.03 mag. No statistically significant periodic signal is present in the data.

Assuming that the average uncertainty in the photometric data represents an upper limit of stellar surface covered by a hypothetical spot (f=0.02) we can estimate after Hatzes (2002) that the resulting RV amplitude might be $\approx$17 m s$^{-1}$. The observer RV semi-amplitude of 26.7$\pm$1.6 is greater, which, together with a lack of RV-BIS correlation and non-zero eccentricity, allows us to reject the spot hypothesis. We note, however, that these two values are rather close.

\subsection {HD 236555 and BD+56 2957 two new spectroscopic  binaries }

For another two stars we obtained first 
Keplerian solutions using HET/HRS data:
BD+56 2957  
and
HD 236555. 
Given the small number of available RV epochs and the time-span of observations shorter than the expected orbital periods, we consider presented solutions preliminary.
We used the Data Analysis Center for Exoplanets (DACE) RV  analysis web module for this purpose. 
 The module utilises the MCMC algorithm described in D\'iaz et al. (2014, 2016) and the Keplerian model initial conditions as in Delisle et al. (2016). 
%


BD+56 2957 was not included in the final PTPS sample because in Zieli\'nski et al. (2012) its atmospheric parameters were not determined. Adamczyk et al. (2015) showed that this is a T$_{eff}$=4000 $\pm$ 250 K, logg=2.0$\pm$0.5 thin galactic disk giant with 
M/M$_{\odot}$=1.44$\pm$ 0.31, 
R/R$_{\odot}$=22.55$\pm$12.01, 
log(L/L$_{\odot}$)=2.17$\pm$0.25.
It hosts, according to our preliminary analysis based on 9 epochs of HET/HRS data only, an m sin{\it i}=1.36 M$_{\odot}$ companion on an moderately eccentric (e=0.31$\pm$0.01) 6.08$\pm$0.14 au  orbit.

HD 236555 too was not included in the final PTPS sample due to missing atmospheric parameters in Zieli\'nski et al. 2012.  According to Adamczyk et al. (2015) it is a T$_{eff}$=5000 $\pm$ 250 K, logg=2.0$\pm$0.5 giant
with
M/M$_{\odot}$=3.01$\pm$ 0.63, 
R/R$_{\odot}$=24.78$\pm$15.11, 
log(L/L$_{\odot}$)=2.39$\pm$0.36.
The star, a thick galactic disk object, hosts an sin{\it i}=107$\pm$12 M$_{J}$ companion on an  9.21$\pm$1.75 au orbit with e=0.28$\pm$0.14.

Results are presented in Table 5 and in Figures 4 and 5.

\MakeTable{lll}{12.5cm}{Two new spectroscopic binaries detected in the sample}
{\hline
Parameter & TYC 4006 00890 1 & TYC 3663 00838 1 \\
 & BD+56 2957 & HD 236555 \\
\hline
\hline
1 & 2 & 3  \\
\hline
P(days) & 4560 $\pm$ 70            & 5700 $\pm$ 1951\\
T$_{0}$ (MJD) & 56105 $\pm$ 48    & 57928 $\pm$ 3311\\
K (m s$^{-1}$) & 10082 $\pm$ 616        & 26.7 $\pm$ 1.6\\
e & 0.31 $\pm$ 0.01                     & 0.28 $\pm$ 0.14\\
$\omega$ (deg) & 211 $\pm$ 4      & 109 $\pm$ 6 \\
m$_{2}$ sin{\it{i}} (m$_{J}$)& 1470 $\pm$ 37   & 107 $\pm$ 12\\
a (au) & 6.08 $\pm$ 0.14                       & 9.21 $\pm$ 1.75  \\
N$_{obs}$ &   9                           & 17 \\
\hline

}

\begin{figure}[htb!]
\includegraphics[width=1.0\textwidth]{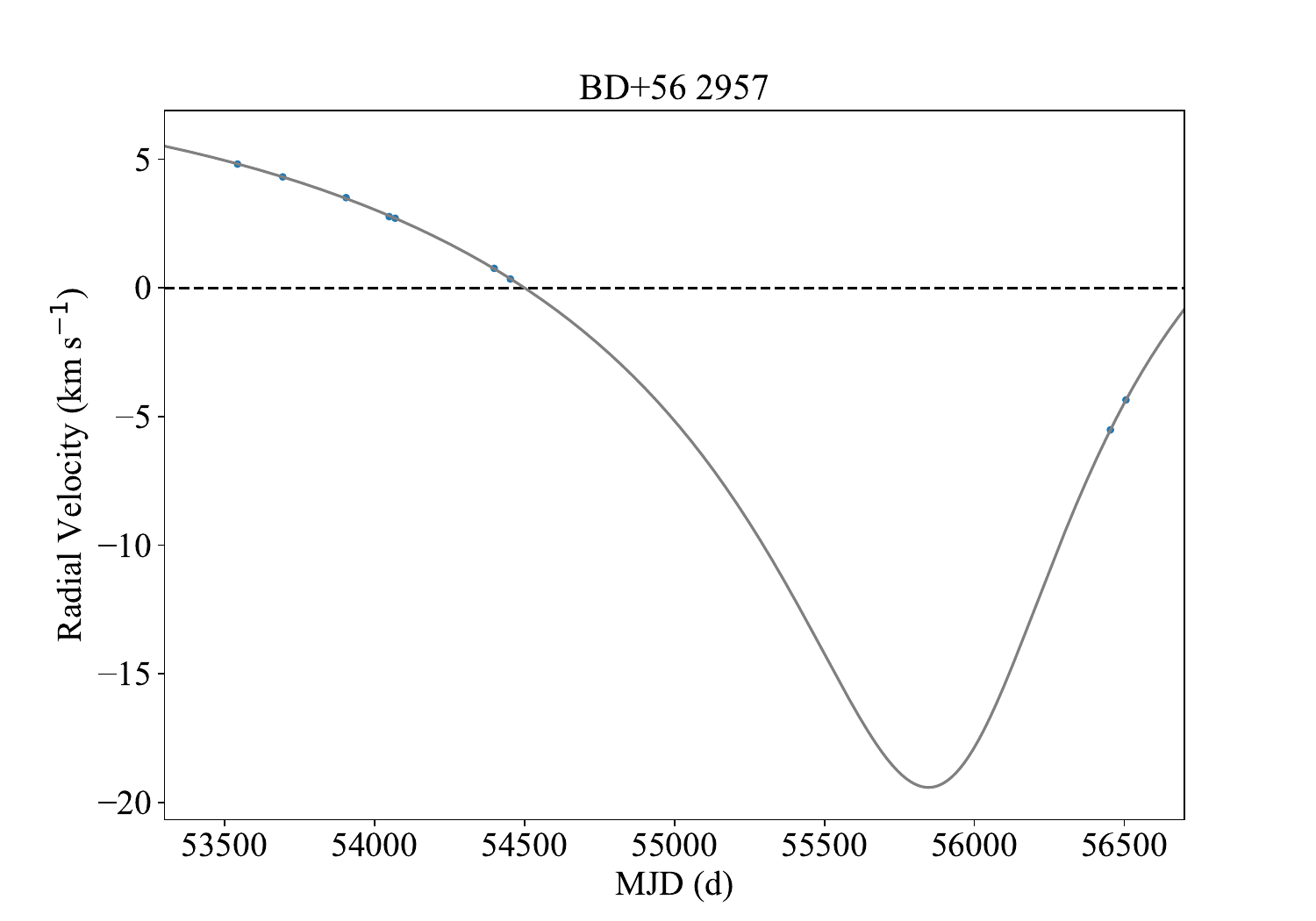}
\FigCap{Preliminary  orbital fit for BD+56 2957}
\end{figure}

\begin{figure}[htb]
\includegraphics[width=1.0\textwidth]{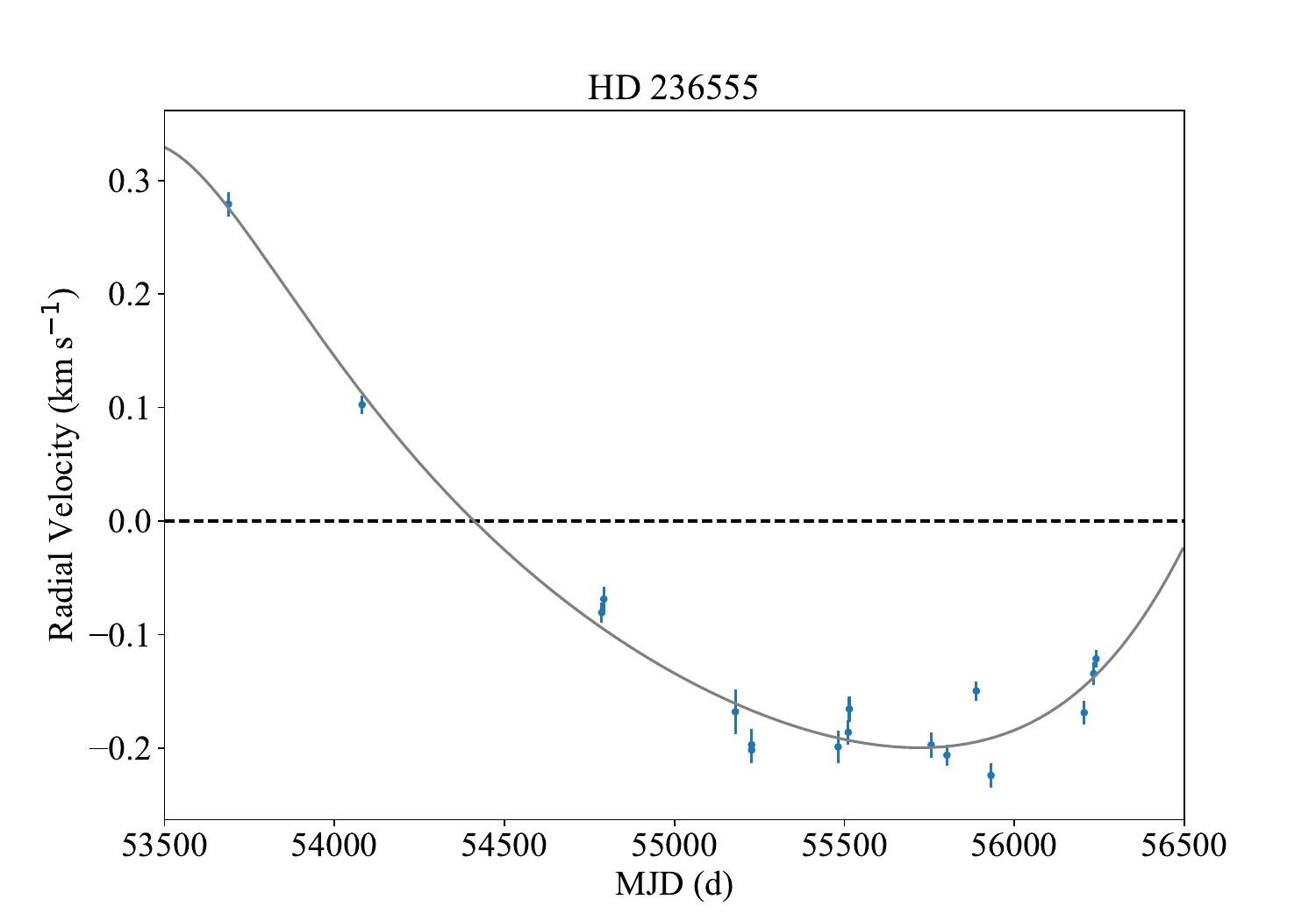}
\FigCap{Preliminary  orbital fit for HD 236555}
\end{figure}

\section{Discussion and summary}

We identified two planetary-mass companions and two binary systems in a sample of 28 stars for which we present precise multi-epoch radial velocities. 

The most intriguing finding of this research is the planetary mass companion to BD+37 3172. The semi-amplitude of radial velocity variations is rather high, about 16 times greater than the mean estimated accuracy of our data. It is also about 5 times greater than the post-fit rms and $\approx$6 times greater than the estimated RV signal due to a hypothetical spot rotating with the star. 

The observed RV variations were present over $\approx$2.5 orbital periods, which makes the detection reliable in the context of 
discussion about the true nature of the observed RV variations in evolved stars in light of such objects as Eltanin ($\gamma$ Dra), where the 702 day period observed between 2003 and 2011 (Hatzes et al. 2018) was not present in more recent (2011-2017) data (op.cit.).

The apparent lack of correlation between radial velocity variations and spectral line bisector displacements suggests that we are dealing with Doppler shifts of spectral lines (we stress here again, however, that in slowly rotating giants the BIS is not a sensitive activity indicator). In this sense the detection appears solid.

The star appears to be among the most massive planetary companions hosts known. With an estimated mass of M/M$_{\odot}$=3.75$\pm$0.86 it is much more massive than the high-mass limit for stellar hosts of  M/M$_{\odot}\approx $2 estimated by Reffert et al. (2015) and  Wolthoff et al. (2022), what makes it very intriguing. Interestingly the orbital separation of 4.65 au fits very well theoretical expectations on the population of exoplanets around the intermediate-mass stars.

The estimated luminosity of log(L/L$_{\odot}$)=2.84$\pm$0.28 places BD+37 3172 near the very intriguing high luminosity evolved 
Long Secondary Period 
stars in the luminosity-period sequence, in line with objects like HD 4760 (Niedzielski et al.  2021). We note however, that the observed periodicity in BD+37 3172 is longer than typical period of LSP - 250-1400 days, and the radial velocity variations amplitude  is much lower  (Wood et al. 2004, Nicholls et al. 2009). Apparent lack of photometric variability also advocates against LSP origin of the observed RV variations in BD+37 3172.

In contrast BD+42 2315 appears to be a regular, roughly solar-mass red giant near the horizontal branch. Its low-mass companion,  the least massive around evolved stars,  appears to orbit the star within the avoidance zone (Villaver and Livio 2007, 2009), close enough to the star to be engulfed in turn of tidal interactions as the host evolves farther. Its orbital separation of a=0.54 au equals to only $\approx$22 R$_{\star}$. The origin of very high eccentricity of the close companion's orbit is therefore unclear.

The detection of a low mass companion to BD+42 2315 should be considered with caution. We note that the modelled radial velocity semi-amplitude K=26.7$\pm$1.6 m s$^{-1}$ is only $\approx$4 times greater that the mean radial velocity uncertainty for this object. It is also comparable to the post-fit rms of $\approx$17 m s$^{-1}$, estimated amplitude of p-mode oscillations, and to the estimated RV signal from a hypothetical spot.
The observed, presumably Keplerian signal is very weak given the precision of our data and much more observations of higher precision  are needed to confirm it.

The two new preliminary Keplerian solutions for BD+56 2957 and BD+42 170 show orbital periods much longer than the time-span of our observations which, together with rather low number of observations makes them uncertain. We note, however, that given $\approx$20$\%$ uncertainty in hosts mass, the companion of  BD+42 170  may be a thick galactic disk brown dwarf.

The high precision, multi-epoch radial velocity data presented in this paper may serve as extension of data-span for other planet searches.

\Acknow{This publication makes use of The Data \& Analysis Center for Exoplanets (DACE), which is a facility based at the University of Geneva (CH) dedicated to extrasolar planets data visualisation, exchange and analysis. DACE is a platform of the Swiss National Centre of Competence in Research (NCCR) PlanetS, federating the Swiss expertise in Exoplanet research. The DACE platform is available at https://dace.unige.ch. This research has made use of the SIMBAD database, operated at CDS, Strasbourg, France. This research has made use of
NASA's Astrophysics Data System. This research made use of Astropy, a community-developed core
Python package for Astronomy (Astropy2013).}


\end{document}